\documentclass{ifacconf}
\usepackage{graphicx}      
\usepackage{natbib}        
\usepackage{amsmath}
\usepackage{url}
\begin{document}
\begin{frontmatter}

\title{Dynamic modeling and simulation of a \\flash clay calciner} 

\thanks{The project has been funded by EUDP in the ECoClay project 64021-7009.}


\author[First]{Nicola Cantisani} 
\author[First,Second]{Jan Lorenz Svensen}
\author[Second]{Ole Fink Hansen} 
\author[First]{John Bagterp Jørgensen}

\address[First]{Department of Applied Mathematics and Computer Science, Technical University of Denmark, DK-2800 Kgs. Lyngby, Denmark}
\address[Second]{FLSmidth A/S, DK-2500 Valby, Denmark}

\begin{abstract}                
We present a novel dynamic model of a flash clay calciner. The model consists of thermophysical properties, reaction kinetics and stoichiometry, transport, mass and energy balances, and algebraic constraints. This gives rise to a system of partial differential-algebraic equations (PDAE). Spatial discretization is performed to convert the PDAEs into a system of differential-algebraic equations (DAE). The model can be used, for example, to perform dynamic simulations with changing inputs, and process design and optimization. Moreover, it can be used to develop model-based control, which is relevant for flexible operation of a clay calcination plant for green cement production.
\end{abstract}


\end{frontmatter}

\section{Introduction}
Cement manufacturing is one of the largest sources of carbon dioxide emissions, accounting for roughly 6\%  \citep{Imbabi:etal:2012}. Accordingly, there is high interest in developing improved production methods that can reduce emissions. Clinker is the main component of cement, and it is produced by burning limestone in the rotary kiln. Around 40\% of the emissions are due to the burning of fossil fuel in the kiln, 50\% are related to the chemical process of calcination of limestone, and the remaining 10\% are indirect emissions. Emissions reduction can be achieved in two ways: 1) by lowering the clinker-to-cement ratio, 2) by substituting fossil fuel with renewable energy. In recent years, calcined kaolinite-rich clay as a clinker substitute has gained a lot of momentum, because of its abundance in nature and its CO$_2$-free calcination process. Substitution of up to 50\% clinker content in cement blends is viable, achieving similar mechanical properties and even improving some aspects of durability  \citep{SCRIVENER201849}. Calcined clay limestone cements are referred to as LC$^3$. 
By electrifying the clay calcination process using renewable energy, emissions reduction of up to a total of 50\% per ton of cement can be achieved. Furthermore, the use of electricity instead of fuel enables better temperature control, and thus higher product quality.

The core of the clay calcination process is the calciner. Because of the intermittent nature of renewable energy sources, it is relevant to be able to dynamically simulate and predict the effect of varying power input and other process conditions (e.g. different clay compositions) in the reactor. A dynamic model of the process is therefore necessary. Ultimately, such a model can unlock the development of model-based control techniques, like model predictive control (MPC), for flexible and optimized process operation.

There is very little literature on dynamic modeling of the clay calcination process.
\cite{Eskelinen2015} present a dynamic model for clay calcination in a multiple hearth furnace. They assume constant heat capacity for the solid phase and the model cannot be solved with a standard solver, but requires a special solution algorithm. 

In this paper, we present a novel complete dynamic model of a flash clay calciner. The model is formulated as a system of partial differential-algebraic equations (PDAE), but translated into differential-algebraic equations (DAE) after spatial discretization. The model is based on first principles, i.e. mass and energy balances. We use a rigorous approach that incorporates thermodynamic functions as algebraic constraints. This allows us to handle complicated (non-constant) expressions of the heat capacity, and have the state variables, such as temperature and pressure, as algebraic variables. This technique makes the model not only more realistic, but easy to implement as a standard DAE system. Moreover, our formulation in blocks allows easy modifications, if needed.

The paper is structured as follows. Section \ref{sec:process_description} provides a short overview and a description of the clay calcination process. Section \ref{sec:model} presents the dynamic model of the calciner, as a system of PDAEs. Section \ref{sec:discretization} presents a spatial discretization of the model, in order to translate it into a system of DAEs. Section \ref{sec:simulation} presents some simulation results of the model. Section \ref{sec:conclusion} concludes the paper.

\section{Process description}\label{sec:process_description}
\begin{figure*}[tb]
    \centering
    \includegraphics[width=0.76\textwidth]{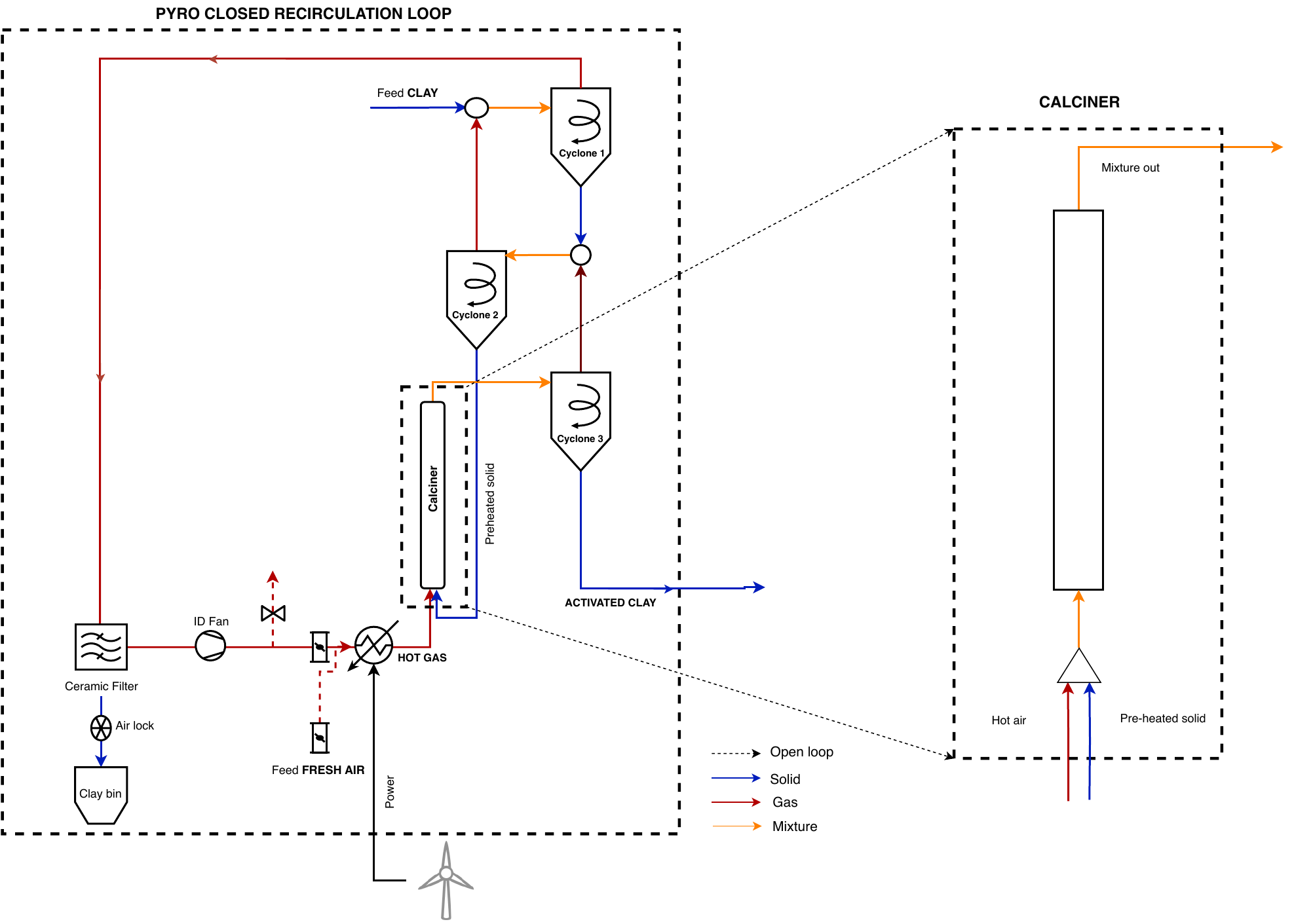}
    \caption{Clay pyro-activation loop. The process consists of a pre-heating zone with 2 cyclones, a calciner, an electric hot gas generator, a separation cyclone, and a gas recirculation loop. The right part shows a zoom of the calciner.}
    \label{fig:processdiagram}
\end{figure*}

Fig. \ref{fig:processdiagram} shows a diagram of the clay calcination process that we consider. We hereby provide a short description.

The thermal activation of the clay is performed in a pyro loop. The fresh clay is introduced in the loop after being crushed. The material stream undergoes pre-heating through two cyclones, where a part of the clay already gets calcinated because of the high temperature. The pre-heated solid is then introduced in the calciner, which is the part that we model in this paper. The calciner is a long plug-flow reactor (PFR) where the solid material stream is mixed with the hot gas stream coming from the electric hot gas generator. The hot gas transfers heat to the solid particles, ensuring that all the clay gets calcined. The last cyclone separates the solid product from the gas before leaving the process. The gas can be recirculated in the loop, in order to recover energy. Since the hot gas generator runs on renewable energy and the clay does not release any carbon dioxide, the process is CO$_2$ free.

\section{Flash clay calciner model} \label{sec:model}
The dynamic model consists of several subparts. We present and discuss them in this section, in the following order: 1) chemical model (reaction kinetics and stoichiometry), 2) thermophysical model (enthalpy and volume functions), 3) transport model, 4) mass balance, 5) energy balance, 6) algebraic relations.
A system of PDAEs arises from the combination of all the components of the model. The same modeling technique is used by \cite{ROSBO2023108316}.

We model the calciner as a PFR of length $L$ and diameter $d$. The reactor volume is denoted as $V_{tot}$. Furthermore, we specify all the concentrations with respect to the reactor volume, i.e.
\begin{equation}
    c = \frac{n}{V_{tot}}.
\end{equation}
We use $c$ for concentration and $n$ for number of moles.

\subsection{Stoichiometry and kinetics}
The type of clay that we consider is composed of kaolinite and quartz. The main reaction occurring when calcinating clay is dehydroxylation of kaolinite. The reaction leads to the formation of metakaolin and water vapor
\begin{equation}
    \begin{split}
        \mathrm{Al_2O_2 \cdot 2SiO_2 \cdot 2H_2O (s) \rightarrow }& \mathrm{ Al_2O_2 \cdot 2SiO_2 (s)} \\ & \mathrm{ + 2 H_2O (g).}
    \end{split}
\end{equation}
    
The temperature range of the reaction is 450-700 ${}^\circ$C. The reaction is endothermic. Notice that quartz is not involved in the chemical reaction. For simplicity in the notation, we indicate hereafter the species as
\begin{equation}
     \mathrm{AB_2 \rightarrow A + 2 B}
\end{equation}
We model the reaction kinetics as a third-order reaction, with activation energy $E_A=202$ kJ/mol and pre-exponential factor $k_0 = 2.9 \times 10^{15}$ s$^{-1}$ \citep{PTACEK201024}. Hence, the reaction rate is
\begin{equation} \label{eq:reactionrate}
    r = r(c_{AB_2},T_s) = k \, c_{AB_2}^3
\end{equation}
where 
\begin{equation}
    k = k(T_s) = k_0 \exp \left( - \frac{E_A}{R T_s} \right).
\end{equation}
$T_s$ is the temperature of the solid. We use the following indexes throughout the paper: AB$_2$ = kaolinite (solid), A = metakaolin (solid), B = water (gas), air = dry air (gas), Q = quartz (solid). 
The chemical production rate is
\begin{equation}
    R = \nu'r(c)
\end{equation}
where $\nu$ is the stoichiometric matrix.
The concentration vector
and the stochiomatric matrix have the form
\begin{equation}
    c = [c_{AB_2}, c_{A}, c_{B}, c_{air}, c_{Q} ]^T, \quad \ \nu = [-1,1,2,0,0].
\end{equation}
Notice that air and quartz do not participate in the reaction, but they are involved in the mass transport and heat exchange. The hot air is used to heat up the solid, in order for the reaction to occur.

\subsection{Thermodynamic functions}
Expressing energy balances in terms of internal energy makes the model depend on temperature $T$, pressure $P$, and number of moles $n$. By using a thermodynamic library, or by constructing one, we can compute the enthalpy $H$, volume $V$, and (therefore) internal energy $U$, i.e.
\begin{subequations}
    \begin{align}
        V & = V(T,P,n),  \\
        H &= H(T,P,n), \\
        U &= H - PV.
    \end{align}
\end{subequations}
Moreover, notice that the enthalpy and volume functions of a mixture may be computed as
\begin{subequations} \label{eq:mixture}
    \begin{align}
        H(T,P,n) & = \sum_{i} n_i h_i(T,P), \\
        V(T,P,n) & = \sum_{i} n_i v_i(T,P),  
    \end{align}
\end{subequations}
where the index $i$ indicate the $i$-th species of the mixture. $h$ and $v$ indicate molar enthalpy and volume. \eqref{eq:mixture} is only valid when the molar enthalpies do not depend on the molar fractions, like in our case. Since these functions are homogeneous of order 1 with respect to the number of moles, we can divide by the reactor volume and obtain the volumetric quantities. 
We use the $\hat \cdot$ notation to indicate them.
\begin{subequations}
    \begin{align}
        \hat v & = V(T,P,c), \\
        \hat h &= H(T,P,c), \\
        \hat u &= \hat h - P \hat v.
    \end{align}
\end{subequations}
In the same way, we may compute the energy fluxes (indicated by $\Tilde \cdot$)
\begin{subequations}
    \begin{align}
    \Tilde H = H(T, P, N), \\
    \Tilde U = U(T,P,N).
    \end{align}
\end{subequations}
$N$ indicates a molar flux. 


\subsubsection{Solid phase}
The solid phase is the clay material. We assume that the clay is composed of kaolinite, metakaolin, and quartz.
The molar enthalpy of each component $i$ may be evaluated if the expression of the heat capacity $c_{p,i}$ of the material is available.
\begin{subequations}\label{eq:molar_enthalpy}
    \begin{align}
        h_i(T,P) & = h_i(T^0,P^0) + \int_{T^0}^T c_{p,i}(s) ds
    \end{align}
\end{subequations}
We use the enthalpy of formation of the material at standard conditions for the reference state. That is
\begin{equation}
    h_i(T^0=298.15 \text{ K},P^0=1 \text{ bar}) = \Delta H^0_{form,i}.
\end{equation}
Table \ref{tab:heatcap} reports the values for metakaolin and kaolinite. We use the notation
\begin{subequations}
\begin{align}
        H_s & = H_s(T_s,P,n_s), \\
        V_s & = V_s(T_s,P,n_s),
\end{align}
\end{subequations}
to indicate the enthalpy and the volume of the solid material. The number of moles in the solid phase is $n_s = [n_{AB_2}, n_A, n_Q]^T$.

The heat capacity of the solid components may given by the following empirical expression 
\begin{equation}
    c_{P}(T) = k_1 + k_2 T + k_3 T^2 + \frac{k_4}{T} + \frac{k_5}{T^2} + \frac{k_6}{\sqrt{T}},
\end{equation}
with $\{k_i\}_{i=1,...,6}$ being coefficients.
This is only valid within the temperature interval $T_{min} \leq T \leq T_{max}$. Table \ref{tab:heatcap} reports the coefficients for metakaolin and kaolinite \citep{Roohangiz,BALE201635}. The temperature $T$ must be given in K and the heat capacity $c_P$ is in 
$\mathrm{\frac{J}{mol \cdot K}}$.
The molar volume is given by
\begin{equation}
    v(T) = v_1 + v_2 T.
\end{equation}
Table \ref{tab:volume} reports the coefficients for metakaolin and kaolinite \citep{Roohangiz,BALE201635}. \cite{NIST2023} provide the enthalpy and the volume expressions for quartz.

\begin{table}[tb]
    \centering
    \caption{Coefficients for the solid heat capacity and standard enthalpy of formation of metakaolin and kaolinite.}
    \begin{tabular}{|c|c|c|}
    \hline
        \textbf{Coefficients} & \textbf{Metakaolin} & \textbf{Kaolinite}  \\
        \hline
        $k_1$ & 2.294924 $\times 10^{2}$ $\mathrm{\frac{J}{mol \cdot K}}$ & 1.4303 $\times 10^{3}$ $\mathrm{\frac{J}{mol \cdot K}}$ \\
        $k_2$ & 3.68192 $\times 10^{-2}$ & -7.886 $\times 10^{-1}$ \\
        $k_3$ & 0 & 3.034 $\times 10^{-4}$\\
        $k_4$ & 0 & 0 \\
        $k_5$ & -1.456032 $\times 10^{6}$ & 8.334 $\times 10^{6}$ \\
        $k_6$ & 0 & -1.862 $\times 10^{4}$ \\
        \hline
        $T_{min}$ & 298 K & 298 K\\
        $T_{max}$ & 1800 K & 700 K \\
        \hline
        $\Delta H^0_{form}$ & -3.211 $\times 10^{6}$ $\mathrm{\frac{J}{mol}}$ & -4.11959 $\times 10^{6}$ $\mathrm{\frac{J}{mol}}$ \\
        \hline
    \end{tabular}
    \label{tab:heatcap}
\end{table}
    
\begin{table}[tb]
    \centering
    \caption{Coefficients for the molar volume.}
    \begin{tabular}{|c|c|c|c|}
    \hline
        \textbf{Coefficients} & \textbf{Metakaolin} & \textbf{Kaolinite} \\
        \hline
        $v_1$ & 41.4736 $\mathrm{\frac{m^3}{mol}}$ & 30 $\mathrm{\frac{m^3}{mol}}$\\
        $v_2$ & 3.39116 $\times 10^{-3} \mathrm{\frac{m^3}{mol \cdot K}}$  & 0 $\mathrm{\frac{m^3}{mol \cdot K}}$\\
        \hline
    \end{tabular}
    \label{tab:volume}
\end{table}

\subsubsection{Gas phase}
The gas phase in the clay calcination process consists of dry air and water vapor. We treat air as an ideal gas mixture made of 78\% nitrogen (N$_2$), 21\% oxygen (O$_2$), and 1\% argon (Ar). The molar enthalpy of each component is computed as for the solid \eqref{eq:molar_enthalpy}. We indicate the enthalpy and the volume of the gas mixture with 
\begin{subequations}
\begin{align}
        H_g & = H_g(T_g,P,n_g), \\
        V_g &= V_g(T_g,P,n_g).
\end{align}
\end{subequations}
$T_g$ indicates the temperature of the gas phase. The number of moles in the gas phase is $n_g = [n_{air},n_{B}]^T$.
\cite{NIST2023} provide the enthalpy expressions for nitrogen, oxygen, argon, and water vapor.

The molar volume of a single gas may be computed using the ideal gas law
\begin{equation}
    v_{g,i} =  \frac{R T_g}{P}.
\end{equation}

\subsection{Transport model}
Mass and energy balances in the calciner directly depend on the spatial material flux. We consider advection and diffusion for the flux. The flux depends directly on the flow velocity.

\subsubsection{Velocity} The velocity along the reactor, $v$, may be modeled as a function of the pressure drop, $\Delta P$, along the length, $\Delta z$, using the Darcy-Weisbach equation for turbulent flows \citep{Svensen2023}.
\begin{equation}
    v = v\Big( \frac{\Delta P}{\Delta z} \Big) = \Big( \frac{2}{0.316} \sqrt[4]{\frac{d^5}{\mu \, \rho^3}} \frac{|\Delta P|}{\Delta z} \Big)^{ \frac{4}{7} } \mathrm{sgn} \Big( \frac{\Delta P}{\Delta z} \Big).
\end{equation}
$\rho$ is the density of the mixture, and $\mu$ is the viscosity of the mixture. The equation is only valid for Mach number $<$ 0.2. The density of the mixture may be computed as
\begin{equation}
    \rho = \sum_j M_j c_j.
\end{equation}
$M_j$ and $c_j$ are the molar mass and the concentration of the $j$-th component.
\subsubsection{Viscosity} 
The viscosity of a suspended gas mixture may be computed by the extended Einstein equation of viscosity \citep{Viscosity2006,Svensen2023}.
\begin{equation}
    \mu = \mu_g \frac{1 + \hat v_s/2}{1 - 2 \hat v_s}.
\end{equation}
$\mu_g$ is the viscosity of the gas phase. The viscosity of a mixed gas may be given as 
\begin{subequations}
\begin{align}
\mu_g & =\sum_i \frac{x_i \mu_{g, i}}{\sum_j x_j \phi_{i j}}, \\
\phi_{i j} & =\left(1+\sqrt{\frac{\mu_{g, i}}{\mu_{g, j}}} \sqrt[4]{\frac{M_j}{M_i}}\right)^2\left(2 \sqrt{2} \sqrt{1+\frac{M_i}{M_j}}\right)^{-1},
\end{align}
\end{subequations}
where $\mu_{g,i}$ is the viscosity of a single gas \citep{Wilke:2004}. Its temperature dependence may be expressed as in \citep{Sutherland:1893}
\begin{equation}
    \mu_{g,i} = \mu_0 \Big(\frac{T}{T_0} \Big)^\frac{3}{2} \frac{T_0 + S_\mu}{T + S_\mu},
\end{equation}
where $S_\mu$ can be calibrated given two measures of viscosity.
\subsubsection{Advection and diffusion}
The molar flux of the mixture, $N$, is modeled as the sum of advection, $N_a$, and Fick's diffusion, $N_d$.
\begin{equation}
    N = N_a + N_d,
\end{equation}
where
\begin{subequations}
    \begin{align}
         N_a & = v \cdot c, \\
         N_d & = -D  \odot \partial_z c.
    \end{align}
\end{subequations}
$D$ are the diffusion coefficients, and $\odot$ is the element-wise product. The flux vector has the form
\begin{equation}
   N = [ N_{AB_2}, N_{A}, N_{B}, N_{air}, N_{Q} ]^T.
\end{equation}
\subsection{Mass balance} 
The mass balance in the calciner reads (PFR model in \cite{nielsen2023a})
\begin{equation}\label{eq:massbalance}
    \partial_t c = - \partial_z N + R.
\end{equation}
The PDE \eqref{eq:massbalance} describes the dynamics of the concentrations $c(t,z)$ in time and space. The variable $z$ is the calciner length direction. $N(t,z)$ are the fluxes and $R(c)$ is the chemical production rate. 
\subsection{Energy balance}
Because two different phases (solid and gas) are interacting and exchanging energy in the calciner, we need to keep track of the temperature of both of them. We hereby derive an energy balance for the solid and the gas phases.

Let us consider a control volume $\Delta V = A \Delta z$. Let us assume that the solid particles are well mixed with the gas, and that they have identical shape and size. We assume that each solid particle is a perfect ball of radius $r_b$.
The accumulated energy \textit{in the solid phase} in the control volume, during the time interval $\Delta t$, is
\begin{equation}\label{eq:energy1}
\begin{split}
    \Delta U_s =& A \, N_s(t,z) h_s(T_s,P) \Delta t - A N_s(t,z+\Delta z) \cdot \\ & h_s(T_s,P) \Delta t + J_{sg} \Delta t - Q_{amb}\Delta t.
\end{split}
\end{equation}
$J_{sg}$ is the heat transfer rate between solid and gas phases, $Q_{amb}$ is the heat loss to the ambient. We can express the heat transfer between the phases as
\begin{equation} \label{eq:sg}
   J_{sg} = k_{sg} A_{sg} (T_g - T_s).
\end{equation}
$k_{sg}$ and $A_{sg}$ are the solid-to-gas heat transfer coefficient and the transfer area, respectively.
Because of the assumption on the solid particles, we can compute the transfer area between the solid and the gas as
\begin{equation}
    A_{sg} = n_{ball} A_{ball} = \frac{V_s}{V_{ball}}A_{ball} = \frac{3 V_s}{r_b}.
\end{equation}
$n_{ball}$ is the number of balls in the volume, $V_{ball}$ is the volume of a ball and $A_{ball}$ is the area of a ball.

Now inserting \eqref{eq:sg} in \eqref{eq:energy1}, and dividing by $\Delta V$ and $\Delta t$, we get 
\begin{equation}
    \begin{split}
         \frac{\Delta U_s}{\Delta t \Delta V} = & - \frac{N_s(t,z+\Delta z) h_s(T_s,P) - N_s(t,z) h_s(T_s,P)}{\Delta z} \\
    & + k_{sg} \frac{3 V_s}{r_b \Delta V} (T_g - T_s) - \frac{Q_{amb,s}}{\Delta V} 
    \end{split}
\end{equation}

Using volumetric quantities and letting $\Delta t \rightarrow 0$ and $\Delta z \rightarrow 0$, the following PDE arises
\begin{equation}
    \frac{\partial \hat u_s}{\partial t} = - \frac{\partial \tilde H_s}{\partial z}  + k_{sg} \frac{3 \hat v_s}{r_b} (T_g - T_s) - \hat Q_{amb,s}.
\end{equation}

The same derivation can be repeated for the gas phase. In compact notation, the following set of PDE describes the (volumetric) energy balance of the solid and gas phases in the calciner
\begin{subequations}\label{eq:energybalance}
    \begin{align}
        \partial_t \hat u_s & = - \partial_z \Tilde H_s  + \hat J_{sg} + \hat Q_{amb,s}, \\
        \partial_t \hat u_g & = - \partial_z \Tilde H_g - \hat J_{sg} + \hat Q_{amb,g}.
    \end{align}
\end{subequations}
The volumetric heat transfer between solid and gas is 
\begin{equation}
    \hat J_{sg} = k_{sg} \frac{3 \hat v_s}{r_b} (T_g - T_s).
\end{equation}
\subsection{Algebraic relations}
Some extra algebraic equations are needed to solve the system. The volume of solid and gas phases should sum to the reactor volume, i.e.
\begin{equation} \label{eq:volume_alg}
    \hat v_g + \hat v_s - 1 = 0.
\end{equation}
Moreover, the differential variables $\hat u_s$ and $\hat u_g$ should match the quantities computed via the thermodynamic functions, that is
\begin{subequations} \label{eq:intenergy_alg}
    \begin{align}
         & U(T_{s},P,c_{s}) - \hat u_{s} =0, \\
         & U(T_{g},P,c_{g}) - \hat u_{g} =0.
    \end{align}
\end{subequations}
$c_{s} = [c_{A},c_{AB_2},c_Q]^T$ are the concentrations in the solid phase, and $c_g = [c_{air},c_B]^T$ are the concentrations in the gas phase.

\subsection{Summary}
The PDAE model consists of the conservation equations \eqref{eq:massbalance} and \eqref{eq:energybalance}, and the algebraic relations \eqref{eq:volume_alg} and \eqref{eq:intenergy_alg}.

\section{PDE spatial discretization} \label{sec:discretization}
\begin{figure}[tb]
    \centering
    \includegraphics[width=0.48\textwidth]{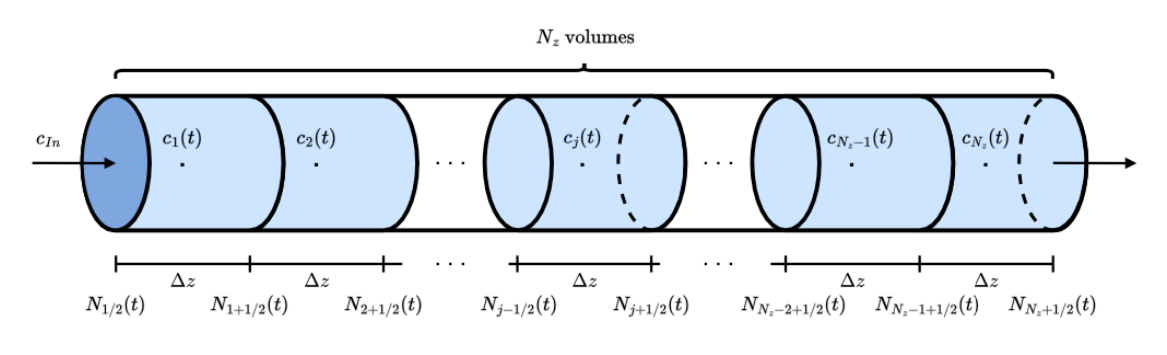}
    \caption{Finite volume discretization of the calciner.}
    \label{fig:finitevolume}
\end{figure}
The mass and energy balances for the calciner resulted in a set of partial differential equations in space and time, \eqref{eq:massbalance} and \eqref{eq:energybalance}. In order to be solved and simulated, we perform spatial discretization by dividing the calciner into finite volumes along the length $z$. We apply central difference approximation to evaluate the derivatives at the center of each cell. Let us consider $N_z$ finite volumes (see Figure \ref{fig:finitevolume}). The fluxes at the cell interfaces are
\begin{equation}
    N_{i+1 / 2}(t) =v_{i+1/2}(t) c_i(t)-D \odot \frac{c_{i+1}(t)-c_i(t)}{\Delta z},
\end{equation}
where the velocities are such that
\begin{equation}
    v_{i+1/2} = v\Big(\frac{P_{i+1}(t)-P_{i}(t)}{\Delta z}\Big),
\end{equation}
for $i \in\left\{1,2, \ldots, N_z-1\right\}$.
At the first and the last cell interface we have
\begin{subequations}
    \begin{align}
        N_{1/2}(t) & = v_{1/2}(t) c_{in},\\
        N_{N_z+1/2}(t) & = v_{N_z+1/2}(t) c_{N_z}(t).
    \end{align}
\end{subequations}
$c_{in}$ are the inlet concentrations.
The velocities at the inlet and outlet of the calciner, namely $v_{1/2}(t)$ and $v_{N_z+1/2}(t)$, depend on the pressures before the inlet and after the outlet, $P_{in}$ and $P_{out}$, respectively.
\begin{subequations}
\begin{align}
    v_{1/2}(t) &= v\Big(\frac{P_1(t)-P_{in}}{\Delta z}\Big), \\
    v_{N_z+1/2}(t) &= v\Big(\frac{P_{out}-P_{N_z}(t)}{\Delta z} \Big).
\end{align}
\end{subequations}
$P_{in}$ and $P_{out}$ are specified as parameters.

The spatial discretization of the mass balance equation results in $N_z$ ordinary differential equations (ODE)
\begin{equation} \label{eq:massODE}
    \frac{d c_i}{d t}(t)=-\frac{N_{i+1 / 2}(t)-N_{i-1 / 2}(t)}{\Delta z}+R\left(c_i(t)\right),
\end{equation}
for $i \in\left\{1,2, \ldots, N_z\right\}$. Notice that $c_i(t)$ is a vector with the 5 chemical components. Therefore the mass balance results in $5 \cdot N_z$ equations.

In the same way, the energy balances are also discretized in space
\begin{subequations} \label{eq:energyODE}
\begin{align}
     \frac{d \hat u_{s,i}}{dt}(t) & = - \frac{\tilde H_{s,i+1/2}(t)-\tilde H_{s,i-1/2}(t)}{\Delta z} + \hat J_{sg,i}(t) + \hat Q_{amb,s}, \\
 \frac{d \hat u_{g,i}}{dt}(t) & = - \frac{\tilde H_{g,i+1/2}(t)-\tilde H_{g,i-1/2}(t)}{\Delta z} - \hat J_{sg,i}(t) + \hat Q_{amb,g}.
\end{align}
\end{subequations}
The enthalpy fluxes of the solid material at the cell interfaces are
\begin{subequations}
    \begin{align}
    \tilde H_{s,1/2}(t) &= H(T_{s,in},P_{in},N_{s,1/2}(t)), \\
\tilde H_{s,i+1/2}(t) &= H(T_{s,i}(t),P_i(t),N_{s,i+1/2}(t)),
    \end{align}
\end{subequations}
for $i \in\left\{1,2, \ldots, N_z\right\}$.
The enthalpy fluxes of the gas at the cell interfaces are
\begin{subequations}
    \begin{align}
    \tilde H_{g,1/2}(t) &= H(T_{g,in},P_{in},N_{g,1/2}(t)), \\
\tilde H_{g,i+1/2}(t) &= H(T_{g,i}(t),P_i(t),N_{g,i+1/2}(t)),
    \end{align}
\end{subequations}
for $i \in\left\{1,2, \ldots, N_z\right\}$. The inlet temperatures of the solid and gas, $T_{s,in}$ and $T_{g,in}$, are manipulated variables.

The heat transfer term between solid and gas is
\begin{equation}
    \hat J_{sg,i}(t) = k_{sg} \frac{3 \hat v_{s,i}(t)}{r_b} (T_{g,i}(t) - T_{s,i}(t)).
\end{equation}
for $i \in\left\{0,1, \ldots, N_z\right\}$.

Finally, the algebraic equations are 
\begin{subequations} \label{eq:algebraic}
    \begin{align}
        & U(T_{s,i}(t),P_i(t),c_{s,i}(t)) - \hat u_{s,i}(t) =0,\\
        & U(T_{g,i}(t),P_i(t),c_{g,i}(t)) - \hat u_{g,i}(t) =0,\\
         \begin{split}
            & V(T_{s,i}(t),P_i(t),c_{s,i}(t)) +  V(T_{g,i}(t),P_i(t),c_{g,i}(t)) \\ & \qquad \qquad \qquad \qquad \qquad \qquad \qquad \qquad \quad - 1 = 0, 
        \end{split} 
    \end{align}
\end{subequations}
for $i \in\left\{0,1, \ldots, N_z\right\}$.

The total discretized ODE model consists of \eqref{eq:massODE}, \eqref{eq:energyODE}, and \eqref{eq:algebraic}. This results in $10 \cdot N_z$ equations. The differential variables $x$ and the algebraic variables $y$ are
\begin{equation}
    x = [c_i; \hat u_{s,i}; \hat u_{g,i}], \quad y = [T_{s,i}; T_{g,i}; P_{i} ].
\end{equation}
for $i \in\left\{1,2, \ldots, N_z\right\}$.
The manipulated variables $u$ and the disturbances $d$ are
\begin{equation}
    u = [ c_{in};  T_{s,in}; T_{g,in} ], \quad  d = [Q_{amb,s};Q_{amb,g}].
\end{equation}
We remark that the concentrations $c_i$ and $c_{in}$ are vectors.

\section{Simulation results} \label{sec:simulation}
The full DAE system may be simulated easily in Matlab by using \verb|ode15s|. The differential and algebraic variables are specified by giving a mass matrix as an input to the function. 
We simulate using the following manipulated variables
\begin{subequations}
    \begin{align}
        c_{in} & = \Big[0.15;0.31;3.74;5.81;0.79 \Big] \mathrm{ mol/m^3}, \\
        T_{s,in} &= 657.15 \text{ K}, \quad T_{g,in} = 1261.15 \text{ K}.
    \end{align}
\end{subequations}

We use $N_z=20$ and a pressure drop of 600 Pa. We neglect the heat loss to the ambient (adiabatic reactor). The diffusion coefficients are all set to 0.1. The initial concentrations are $[0.1;0.1;0.1;19.65;0.1]$ $\mathrm{ mol/m^3}$ in all the cells. The initial temperature of the solid and the gas is 600 K. 

\begin{figure}[tb]
    \centering
    \includegraphics[width=0.48\textwidth]{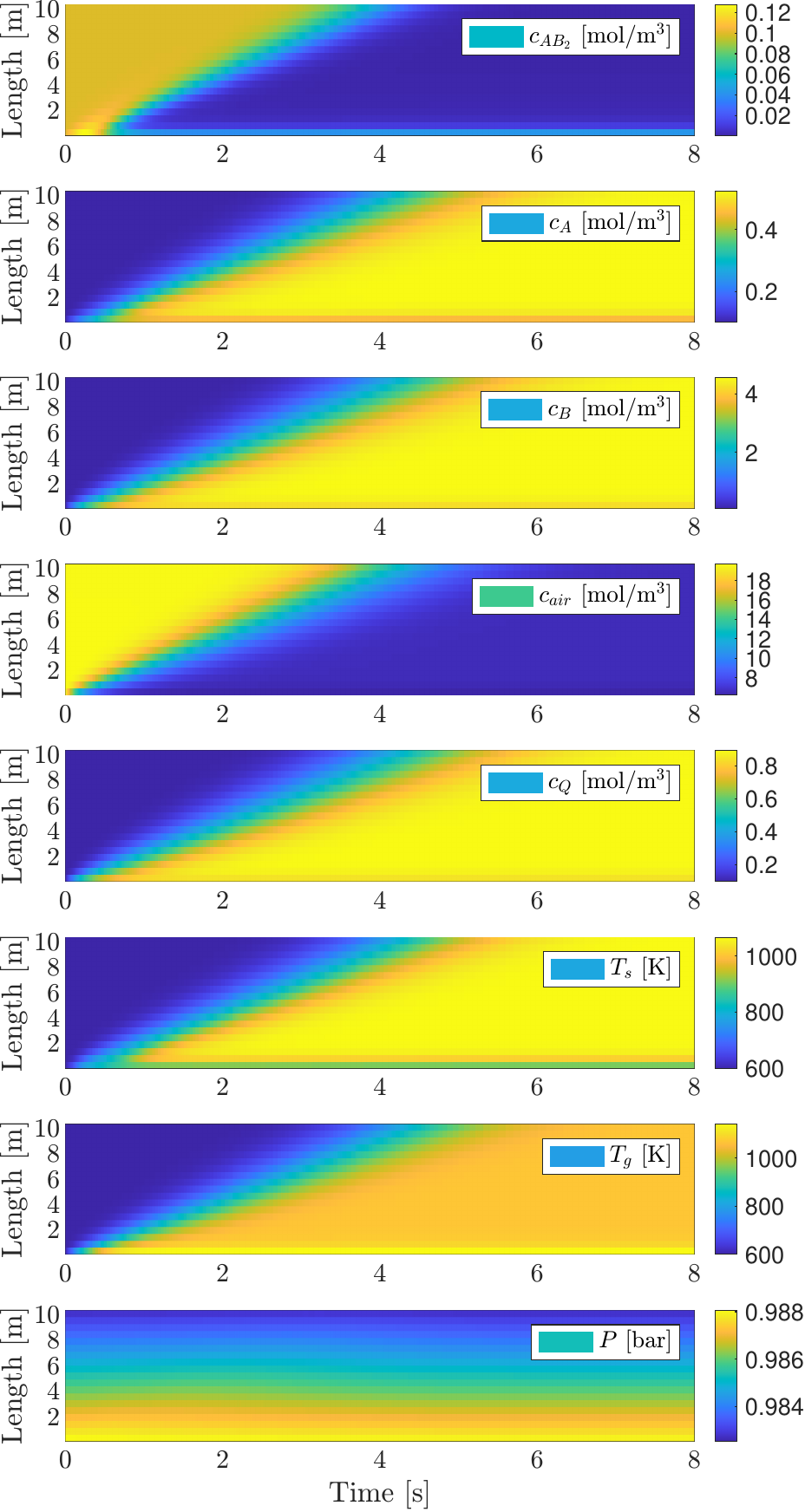}
    \caption{States in time and space: concentrations, temperature of the solid and the gas, pressure.}
    \label{fig:conc3d}
\end{figure}

\begin{figure}[tb]
    \centering
    \includegraphics[width=0.46\textwidth]{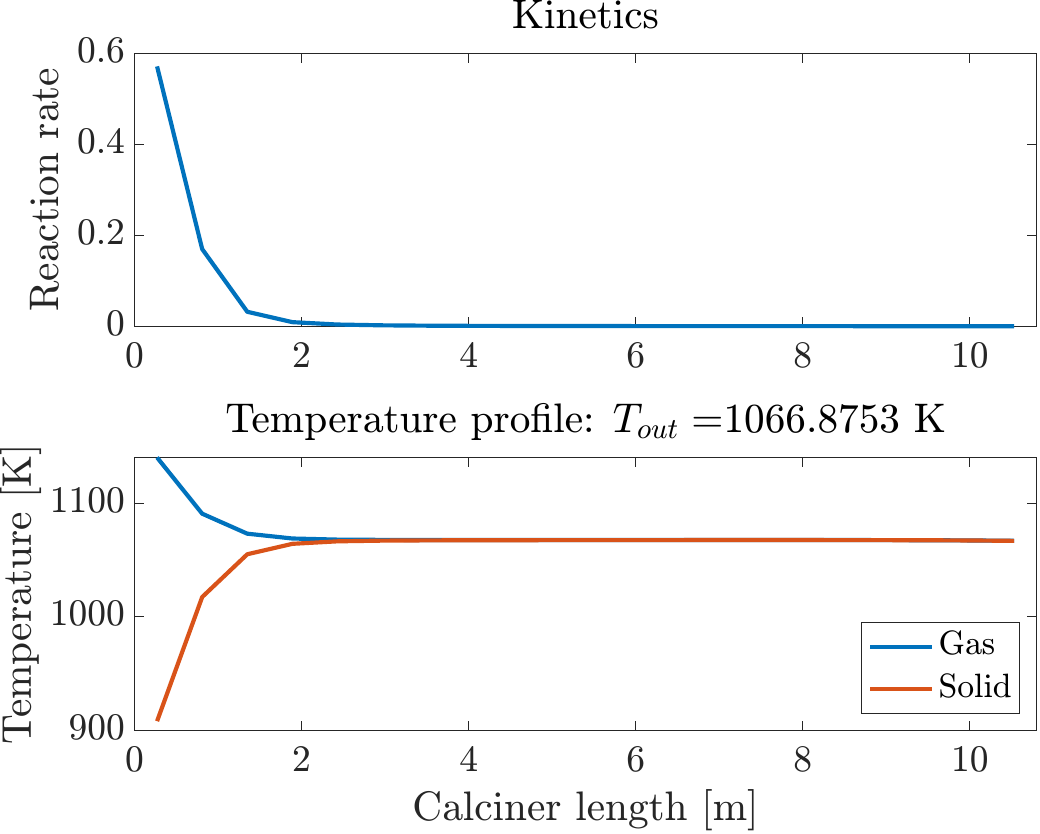}
    \caption{Steady state: reaction kinetics and gas-solid temperature profile along the calciner.}
    \label{fig:temp_Steady}
\end{figure}

Fig. \ref{fig:conc3d} displays 3D plots in time and space of the states, i.e. concentrations, temperature of the solid and the gas, and pressure. Steady state is reached after few seconds, because of the fast dynamics. Fig. \ref{fig:temp_Steady} shows the reaction rate and the solid and gas temperature profiles along the calciner at steady state. The reaction and the heat transfer happen mostly at the beginning of the calciner.

\section{Conclusion} \label{sec:conclusion}
A dynamic model for a clay calciner is presented and discussed. The model consists of several building blocks, which are formulated in a way that allows easy modification if necessary. For example, extra side reactions can be added, and different transport models can be investigated, with very little effort in the implementation. The rigorous thermodynamic functions incorporation allows realistic evaluation of heat transfer phenomena at changing conditions. The paper also presents simulation results, showing a dynamic simulation and additional steady state results.

\bibliography{ifacconf}             
                                                   







\end{document}